\documentstyle[12pt,aasms,epsf]{article} 

\newcommand{\asec}{$^{\prime \prime}$}
\newcommand{\rf}{\reference}
\newcommand{\nh}{$N_H$}
\newcommand{\nhg}{$N_{H,gal}$}
\newcommand{\um}{~$\mu$m}
\newcommand{\gam}{$\Gamma$}
\newcommand{\pin}{photon index}
\newcommand{\PIN}{Photon Index}
\newcommand{\kmsmpc}{km~s$^{-1}$~Mpc$^{-1}$}
\newcommand{\lsx}{$L_{SX}$}
\newcommand{\Sxr}{Soft X--ray}
\newcommand{\sxr}{soft X--ray}
\newcommand{\sxrs}{soft X--rays}

\newcommand{\hxr}{hard X--ray}
\newcommand{\hxrs}{hard X--rays}
\newcommand{\SS}[1]{\S~\ref{#1}}
\newcommand{\mc}[2]{\multicolumn{#1}{#2}}
\newcommand{\los}{line of sight}

\newcommand{\gapprox}{\mbox{$\stackrel {>}{_{\sim}}$}}   

\begin{document} 

\title{SOFT X--RAY PROPERTIES OF SEYFERT GALAXIES IN THE ROSAT ALL--SKY SURVEY}

\author{Brian Rush\altaffilmark{1} and Matthew A. Malkan} \affil{Department of
Physics and Astronomy, University of California, Los Angeles, CA 90095--1562;
rush,malkan@astro.ucla.edu}

\altaffiltext{1}{Current Address: The Carnegie Observatories, 813 Santa Barbara
St., Pasadena, CA 91101--1292}


\and

\author{Henner H. Fink and Wolfgang Voges} \affil{Max Plank Institut f\"{u}r
Extraterrestrische Physik, Giessenbachstrasse, D--85740 Garching bei
M\"{u}nchen, Germany; hhf,whv@rosat.mpe-garching.mpg.de}

\begin{abstract} 

We present the results of ROSAT All--Sky Survey observations of Seyfert and
IR--luminous galaxies from the Extended 12\um\ Galaxy Sample and the
optically--selected CfA Sample. Detections are available for 80\% (44/55) of
the Seyfert~1s and 34\% (23/67) of the Seyfert~2s in the 12\um\ sample, and for
76\% (26/34) of the Seyfert~1s and 38\% (6/16) of the Seyfert~2s in the CfA
sample.  Roughly half of the Seyferts (mostly Seyfert~1s) have been fitted to
an absorbed power--law model, yielding an average \pin\ of \gam=2.26$\pm$0.11
for 43 Seyfert~1s and \gam=2.45$\pm$0.18 for 10 Seyfert~2s, with both types
having a median value of 2.3.


The \sxr\ luminosity correlates with the 12\um\ luminosity, with Seyfert~1s
having relatively more \sxr\ emission than Seyfert~2s of similar mid--infrared
luminosities, by a factor of $1.6\pm0.3$. Several physical interpretations of
these results are discussed, including the standard unified model for Seyfert
galaxies.  Infrared--luminous non--Seyferts are shown to have similar
distributions of \sxr\ luminosity and X--ray--to--IR slope as Seyfert~2s,
suggesting that some of them may harbor obscured active nuclei (as has already
been shown to be true for several objects) and/or that the soft X--rays from
some Seyferts~2s may be non--nuclear.

A \sxr\ luminosity function (XLF) is calculated for the 12\um\ sample, which is
well described by a single power--law with a slope of --1.75. The normalization
of this XLF agrees well with that of a \hxr\ selected sample.  Several of our
results, related to the XLF and the X--ray--to--IR relation are shown to be
consistent with the \hxr\ observations of the 12\um\ sample by Barcons et al.

\end{abstract} 


\section{Introduction} \label{intro} 


The enormous interest in ROSAT's potential for the study of active galactic
nuclei (AGN) is indicated by the fact that roughly one--third of pointed
observation time is spent on these objects. Initial analysis of the ROSAT
All--Sky Survey (hereafter RASS; Voges 1992, Brinkmann 1992) shows that ROSAT
sees the vast majority of catalogued Seyfert~1 galaxies. The average power--law
photon index of ROSAT AGN (QSOs and Seyferts) is $\sim2.4\pm1.0$ (Bade et
al. 1995), similar to that found by Hasinger et al. (1991) in their analysis of
deep field observations. The values measured by ROSAT are typically much
steeper than the canonical \hxr\ (2---20~keV) value of $\sim$1.7 measured by
HEAO--1 and the Einstein SSS, and sometimes even steeper than the lower
resolution, \sxr\ (0.2---4.5~keV) slopes found by the Einstein IPC (e.g., the
average IPC \pin\ for radio--quiet quasars of 2.0 (Wilkes \& Elvis 1987); also
see the review by Mushotzky, Done, \& Pounds 1993, for summaries of these
earlier, higher--energy missions). However, certain types of Seyferts, in
particular ``ultra--soft" X--ray--emitting Seyferts, narrow--line Seyfert~1
galaxies (NLS1s) and previously published Seyfert~2s typically have steeper
ROSAT spectra. As discussed in this paper, the X--ray \pin\ is a critical
observational parameter, with competing physical models for AGN predicting
different observed values.

In the current paper, we present of ROSAT All--Sky Survey (RASS) data for two
samples of Seyfert galaxies: those in the mid--infrared selected 12--Micron
sample (Rush, Malkan, \& Spinoglio 1993, hereafter RMS) and the optically
selected CfA Sample (Huchra \& Burg 1992). As discussed in the next section,
these two samples were chosen because they are large and well--defined, and,
being bright and nearby, they will complement studies of higher--redshift
AGN. In \SS{targets} we discuss the selection of targets and the observations;
in \SS{results} we present the results of the X--ray observations, and make
comparisons with observations at longer wavelengths; in
\SS{interp} various physical interpretations of the results are discussed, with
particular emphasis given to the unified model of Seyfert galaxies; in
\SS{xlfs} the Seyfert galaxy \sxr\ luminosity function is calculated from the
12\um\ sample; and in \SS{summary} we summarize the results.



\section{Target Selection and Observations} \label{targets} 

\subsection{Advantages of 12--Micron Selection}
\label{targets_selection} 

All of the Seyfert galaxies for which we have obtained RASS data were selected
from the Extended 12 Micron Galaxy Sample (RMS) and the optically selected CfA
Seyfert Galaxy Sample (Huchra \& Burg 1992). These galaxies are ideally suited
to expand our knowledge of the X--ray properties of Seyferts for several
reasons. First, the 12\um\ sample is the most complete and unbiased large
sample of bright AGNs compiled to date (RMS). Spinoglio \& Malkan (1989) showed
that the wide range of types of Seyfert galaxies emit an approximately constant
fraction of their bolometric luminosity in the mid--IR near 12\um\ (and
Spinoglio et al. 1995 showed the same to be true for non--Seyfert spiral
galaxies with a lower constant of proportionality). Thus selecting at 12\um\ is
a good approximation to selecting a sample based on bolometric flux, and will
therefore result in a sample which has an amount of Seyfert~1s and Seyfert~2s
which accurately reflect the relative numbers of such objects in the local
universe. In contrast, the more common far--IR (e.g., 60\um) selection
technique is biased towards reddened objects which emit a large fraction of
their energy at long wavelengths as a result of dust--reprocessing of
higher--energy photons, while optical/UV samples will be biased towards less
dusty objects. The unbiased nature of the 12\um\ sample will allow us to derive
reliable statistical results applicable to all Seyferts as a class. Although
the CfA sample was selected at optical wavelengths and is suspected of
suffering from some selection affects (Persic et al. 1989; Huchra \& Burg
1992), it also is better than most other samples for studying Seyfert galaxies
(as has been done extensively with Einstein IPC data, by Kruper, Urry, \&
Canizares (1990)).  This is because of the thorough and well--defined nature of
this sample, as the CfA Seyfert sub--sample was defined by complete
spectroscopic observations of the entire CfA sample, the latter containing
every object with $m_{pg}$ brighter than 14.5 in an area covering most of the
northern hemisphere (over 2000 objects in total; Huchra \& Burg 1992).

Second, the 12\um--selected galaxies are qualitatively different from those AGN
analyzed previously in the X--rays. Observing targets such as these is
necessary to determine the {\it range\/} of X--ray properties among AGN, from
the highest to the lowest luminosities, and across the full range of
multiwavelength parameter space. For example, Moran et al. (1992) pointed out
that most of the Seyfert~2s popular with previous observers have polarized
broad lines, and are restricted to those with relatively strong radio
luminosities and UV excesses (e.g., as found by the Markarian surveys).
Compared to these Markarian Seyfert~2s (many of which were observed but not
detected by Ginga---Awaki 1993), the Seyfert~2s we observed have redder
optical/infrared colors, weaker and smaller radio sources, larger starlight
fractions, and steeper Balmer decrements---more representative of the entire
class of Seyfert~2s. Similarly, our Seyfert~1s span over 4 orders of magnitude
in \sxr\ luminosity, from below 10$^{42}~erg~s^{-1}$ to above
10$^{46}~erg~s^{-1}$. This will allow us to determine whether the X-rays are
relatively more dominant, in the low-- and high--luminosity objects (see, for
example, the discussion in \SS{interp}).  Furthermore, many previous samples of
AGN studied at X--ray wavelengths have been been devoted to more distant,
high--luminosity objects (often referred to as quasars, which are presumably
the higher--luminosity, less--resolved counterparts to Seyferts).  The two
samples we discuss contain mostly nearby, bright Seyfert galaxies, which will
complement higher--redshift studies.

The RASS also detected 6 of the 37 IR--luminous non--Seyferts in the 12\um\
sample, and has upper limits for all but two of the others.  These are objects
which do not have the strong and/or broad optical emission lines usually used
to identify a Seyfert, but which do have high far--infrared luminosities
(integrated 8--120\um\ luminosities above $6\cdot10^{44}~erg~s^{-1}$), typical
of Seyferts (RMS).  In \SS{results_lll}, we compare the RASS and
multiwavelength properties of these objects to those of our Seyferts. This is
one way to search for galaxies which may harbor hidden active nuclei. For
example, Boller et al. (1992) compared the RASS to a catalog of 14,708
extragalactic IRAS sources, finding $\sim$250 positional coincidences.  Ten of
these are spirals with non--Seyfert optical classifications, but which reach
X--ray luminosities of a few $10^{43}~erg~s^{-1}$, at least one order of
magnitude higher than any non--Seyferts detected by Einstein (Boller et
al. 1996a). Optical observations of these galaxies show them to have Liner or
HII--region galaxy spectra.  Such X--ray luminous spirals may have buried AGNs,
extreme starbursts, or they may constitute a new class of X--ray emitting
galaxies. Similarly, Ward et al. (1994) obtained optical spectrophotometry of
several very X--ray luminous galaxies in the RASS, finding some to be AGN and
others to be starburst/HII region galaxies. Bade \& Schaeidt (1994) identified
RASS sources on objective prism plates, finding fourteen emission--line
galaxies to have the optical spectra of starburst or Seyfert~2 galaxies. Like
previously known narrow--line X--ray galaxies such as NGC~2110, NGC~2992, and
NGC~5506, some of these objects have Seyfert~2--like optical spectra but
Seyfert~1--like \sxr\ luminosities above 10$^{43}~erg~s^{-1}$.

In addition to finding new Seyferts, comparison with IR--luminous non--Seyferts
can also be used to quantify whether the \sxr\ emission in some Seyferts may be
produced by the normal processes of stellar evolution, as in classic starburst
nuclei like NGC~7714 (Weedman et al. 1981). This is most likely the case for
those Seyferts which emit strongly in the thermal infrared, but relatively
weakly in the X--rays. Even without accurate spectral fitting, the strength of
the X--rays alone can be used to determine whether a non--stellar central
engine is required in the Seyferts and IR--luminous non--Seyferts in our sample
(e.g., as calculated in Boller et al. 1996a; see discussion in
\SS{results_lll}).  However, even when an X--ray emitting central engine is not
required, it may still be present if the object is heavily absorbed in the
ROSAT band.

\subsection{ROSAT All--Sky Survey Observations} \label{targets_obs} 

The data presented in this work were obtained during the ROSAT All--Sky Survey
in 1990---1991, using the Position Sensitive Proportional Counter (PSPC) in the
0.1--2.4~keV energy band. ROSAT was launched from Cape Canaveral on
June~1,~1990 and, after a detailed calibration and verification phase, the
All--Sky Survey started on July~30,~1990 (Brinkmann 1992). The survey was
performed by scanning the sky along great circles in ecliptic longitude,
resulting in images in the form of stripes 360\deg\ long and 2\deg\ wide,
corresponding to the field of view of the XRT+PSPC. The total exposure of a
target ranges from $\sim$500 secs for objects near the ecliptic equator to up
to a few 10$^4$ secs for circumpolar targets. The focal plane instrumentation,
developed and built at the MPE, consists of two redundant PSPCs which are
operated in gas flow mode. The PSPC provides spatial and spectral resolution
over the full field of view of 2\deg\ which varies slightly with photon energy
E.  The energy resolution is $\Delta$E/E = 0.41/$\sqrt{E_{keV}}$. The on--axis
angular resolution is limited by the PSPC to about 25\asec, and the on--axis
effective collecting area, including the PSPC efficiency, is about 220~cm$^2$
at 1~keV (Brinkmann 1992).

We have extracted nearly complete \sxr\ data (count rates or upper--limits, and
in many cases spectral parameters from a power--law fit) from the RASS for the
12\um\ and CfA samples of Seyfert galaxies, and for the IR--luminous
non--Seyferts in the 12\um\ sample.  This includes 54 of the 55 Seyfert~1s,
59/68 Seyfert~2s, and 35/37 IR--luminous non--Seyferts in the 12\um\ sample
(RMS), as well as 47 of the the 50 Seyferts in the optically--selected CfA
Galaxy Sample (Osterbrock \& Martel 1993). Thus, the total number of objects
with ROSAT data presented here is 173 (the brighter $\sim$half of the CfA
Seyferts, including 22 objects reported here, are also in the 12\um\ sample;
this total also includes data from our pointed PSPC observations for the five
objects MKN~1239, NGC~5005, NGC~5135, NGC~424, and NGC~4388---Rush \& Malkan
1996). Actual ROSAT detections are available for 80\% (44/55) of the Seyfert~1s
and 34\% (23/67) of the Seyfert~2s in the 12\um\ sample, and for 76\% (26/34)
of the Seyfert~1s and 38\% (6/16) of the Seyfert~2s in the CfA sample. This is
approximately the same as the Einstein detection rates for the Seyfert~1s, and
slightly higher than that for the Seyfert~2s, as estimated by detections and
upper limits of objects in our sample in the literature (e.g., Kruper et al.
1990; Kriss, Canizares, \& Ricker 1980).

\subsection{Calculations} \label{targets_calcs} 

Applying the standard RASS processing to Photon Event Files, we extracted count
rates in a 300\asec\ circle centered on the IRAS coordinates of each of our
sources (typically accurate to 10---15\asec), subtracting the background as
determined in a concentric annulus with inner and outer radii of 500 and
1500\asec, respectively (with the size of the circle and/or annulus adjusted
when necessary to avoid other sources).  Two--sigma (or better) detections are
given when the background--subtracted source count rate is greater than twice
the square root of the source--plus--background count rate (most detections are
above the 5$\sigma$ level).  Two--sigma upper limits to the count--rate are
given when this is not the case. Uncertainties on the detections are quoted at
1$\sigma$.  When over 50 source counts were available, we fit the spectra to an
absorbed power--law function.  For each object, we calculate the flux in one of
three ways (as flagged in Table~1):
(i) we have an unabsorbed\footnote{We use the terms ``unabsorbed'' and
``absorbed'' to refer to that flux which is measured before and after passing
through \nh\ in the Galaxy, respectively. This distinction does not consider
any affects of internal absorption in the target galaxy. Later (esp. in
\SS{interp}), we use the terms ``intrinsic'' and ``observed'' to distinguish
between the flux emitted from the Seyfert nucleus in the target galaxy and that
which would be measured after passing through absorbing material of any type in
the target galaxy, but still before it is absorbed by \nh\ in our Galaxy.},
monochromatic flux at 1~keV, calculated directly from the model fit;
(ii) there were not enough counts for a model fit (between 30 and 50), but only
to estimate \gam\ from hardness ratios (assuming Galactic \nh, as described
thoroughly in Bade et al. 1995). We then calculate an unabsorbed 0.1--2.4~keV
flux from the RASS count rate, using the energy--to--count conversion factor,
as a function of \gam\ and \nh\ (fixed at \nhg), as given by the XSPEC software
(distributed by the NASA/Goddard Space Flight Center). These conversion factors
are obtained by integrating the ROSAT XRT+PSPC effective area (convolved with
the photon redistribution matrix) over frequency, and are the same as those
used in the model fits in case (i);
(iii) there were not enought counts for a model fit, or to estimate \gam\ from
hardness ratios (less than 30). Thus, we used the median value of \gam=2.3 to
obtain the conversion factor (which contributes errors no worse than a factor
of $\sim$2 even if the actual slopes are extremely steep or flat, and usually
errors of less than $\pm50\%$). The conversion from count rate to unabsorbed
monochromatic flux then was done in the same way as in case (ii).
In either case, we calculate the unabsorbed monochromatic flux or unabsorbed
integrated flux from the other using the power--law model:
$$F_{\mbox{\tiny{0.1--2.4~keV}}}=\int_{0.1}^{2.4}F_\nu\,d\nu$$ with
$$F_\nu\propto\nu^{1-\Gamma}.$$ The luminosity in each case is calculated from
the flux assuming an $H_0$ of 75~\kmsmpc.

In Table~1, we give the following data: (1) object name; (2) object type and
sample (12\um\ and/or CfA); (3--4) the 1950 equatorial coordinates; (5) the
galactic \nh\ value (from Heiles \& Cleary [1979] for the most southern
objects, and from Dickey \& Lockman [1990] or Stark et al. [1992] for the
rest); (6) the RASS count rate (a 2$\sigma$--or--better detection with the
associated 1$\sigma$ uncertainty, or a 2$\sigma$ upper limit); (7) the
unabsorbed, monochromatic flux, $\nu F_{\nu}$, at 1~keV; (8) the photon index,
\gam\ (usually from the fit; cases where this is estimated from hardness ratios
are flagged); and (9) the unabsorbed, integrated 0.1---2.4~keV luminosity.

\section{Results} \label{results} 

\subsection{X--Ray Properties} \label{results_xray} 


Figures~1{\it{a\/}} and~1{\it{b\/}} show histograms of the \pin\ and unabsorbed
\sxr\ luminosity, respectively. The average \pin\ is \gam=2.26$\pm$0.11 for 43
detected Seyfert~1s and \gam=2.45$\pm$0.19 for 10 detected Seyfert~2s
(uncertainties representing one standard deviation). The two classes have
identical median values of \gam=2.29 and 2.28, respectively.  A K--S test
applied to the \pin\ gives only a 19.8\% probability of the Seyfert~1s and~2s
being drawn from different samples, implying that the slight difference in the
average values is not significant.  However, for our measured luminosities, the
median value of 42.75 for 57 Seyfert~1s is nearly an order of magnitude higher
than the value of 41.86 for 24 Seyfert~2s.  A K--S test shows these two
distributions to be drawn from different samples with a 99.98\% probability.
This could simply represent intrinsically distinct distributions of $L_{SX}$
for the two types of Seyfert. Or, the two distributions could be intrinsically
similar, but with the luminosities of the Seyfert~2s systematically suppressed,
for example by scattering and/or obscuration, as discussed in \SS{interp} In
either case, this difference in luminosity is similar to that found by Barcons
et al. (1995), who compared the 5~keV monochromatic luminosities of Seyfert~1s
to those of Seyfert~2s, finding the former to be ~7 times more luminous on the
average, indicating that the luminosity difference between the two classes
spans most well--observed X--ray energies.






In Figure~2, we plot \gam\ versus the unabsorbed 1~keV flux.  We have done the
fits fixing \nh\ to the Galactic value, which provides acceptable fits in most
cases.  Allowing \nh\ to vary as an additional free parameter yields values for
\nh\ consistent with the Galactic value in all but a few cases, implying that
most objects in this plot are not heavily absorbed. We do see, however, that
four of the five flattest objects are also Seyfert~1s with very low fluxes
($\nu F_{1keV}<10^{-11.5}$; the other one, IC~4329, is one of the objects for
which the best--fit \nh\ was significantly larger than the Galactic value).
Since the plotted fluxes are unabsorbed, i.e. not affected by \nh\ in the
Galaxy, these few faint galaxies probably have additional {\it internal\/}
absorption, which would only be obvious with broader energy coverage and many
more photons per spectrum. The effect of this (unaccounted for) extra \sxr\
absorption is to make the observed spectra appear harder, giving directly the
flatter values for \gam. It is also likely that these faint, flat objects are
among the more dust--obscured and reddened Seyfert~1s, since NGC~3227 and
NGC~4235 have been observed to be Seyfert~1.5s (Osterbrock \& Martel 1993), and
MKN~1040 and MKN~6 are listed as Seyfert~1.5s in NED\footnote{the NASA--IPAC
Extragalactic Database}, implying that these objects have relatively faint
broad--line regions. Such absorption effects are likely to affect only the most
internally obscured objects, however, and typically would lead to differences
of only a few to a few tens of a percent in the calculated luminosity (as
verified by fits to the pointed observations in Rush \& Malkan 1996, which were
done both with \nh\ free and with \nh\ set to the Galactic value, and which
included a couple of heavily obscured objects). This uncertainty is much less
than the range of values spanned by Figure~2.  For one of the two extremely
steep objects, NGC~2992, the value of \gam\ measured (4.10) is highly uncertain
because high intrinsic absorption limited the fit to energies above 0.8~keV
(see note to Table~1). The most extreme object (\gam=4.46) is 3C120. The photon
index for this object is calculated from hardness ratios, and is thus less
certain than that for other objects. This value for \gam\ also may be higher
than the true slope if the value for \nhg\ used in the
hardnes--ratio--to--\gam\ calculation (explained in Bade et al. 1995) is higher
than the actual value affecting this observation. This is quite possible, since
the value assumed for $N_H$ is $10.7\cdot10^{20}cm^{-2}$, the second highest of
the 173 objects we observed.  Because of the dependence on \gam\ of the
unabsorbed flux in the model calculations, these two objects also have
uncertain fluxes, further questioning their position on this diagram.


We also investigated any systematic effects of nuclear reddening in the entire
sample, by comparing \gam\ to various indicators of reddening/dust content on
large scales (the 25--60\um, 12--60\um, and 2.2--25\um\ flux ratios, and the
60\um\ luminosity). None of these plots showed any correlation. An inverse
correlation between \gam\ and any of these parameters, would have indicated
that those objects with flatter measured values of gamma are more intrinsically
absorbed, and that the amount of \sxr\ absorption is related to general
dustiness/redness in the target galaxy. However, the lack of any such
correlations implies that the flat--\gam\ objects are not more heavily obscured
on large scales and, thus, that the measured photon indices are intrinsic, {\it
or\/} that any such internal absorption which may be present is not related to
the large--scale reddening in the galaxy.

In addition to altering the fluxes and slopes of detected objects,
absorption---in particular absorption in our Galaxy---could also affect which
objects are detected by ROSAT at all. Figure~3 shows a plot of the unabsorbed
1~keV flux versus the Galactic value of \nh\ towards the source. For objects
above $\log \nu F_{1keV}=-12.3$ there is no significant correlation, but almost
all low--flux objects are detected through low values of \nhg. (This is show by
the small box in the figure, which was positioned specifically to minimize the
effect discussed here, to remove any possible binning affects caused by the
position of a few objects.) This implies that faint objects observed in
directions of high \nhg\ suffer sufficient extinction to have been undetected
in the RASS. Since the distribution on the sky of Seyferts as a function of
brightness is independent of any Galactic properties, we can use this plot to
estimate roughly the number of faint Seyferts in directions of high \nhg\
missed by the RASS. Below $3.75\times10^{20}\mbox{cm}^{-2}$, where extinction
is minimal, we find 23 and 17 12\um--sample Seyferts detected with 1~keV fluxes
above and below $10^{-12.3}$, respectively. However, above
$3.75\times10^{20}\mbox{cm}^{-2}$, we detect 20 Seyferts brighter than
$10^{-12.3}$, and only 7 less luminous, implying that there may be $\sim$8
X--ray--faint objects missing from the latter group. These missing Seyferts
might be accounted for by the objects in our sample at high values of \nh\ with
upper limits in the RASS.



\subsection{Correlations with the Far-Infrared} \label{results_lll} 


Figure~4{\it{a\/}} shows a plot of the unabsorbed, integrated \sxr\ luminosity
from the RASS versus the monochromatic IRAS 12\um\ luminosity. The higher and
lower solid lines represent fits to the 53 Seyfert~1s and 24 Seyfert~2s
detected in either sample, respectively (nearly identical fits are obtained
when fitting only the 12\um\ or CfA samples alone). These fits have slopes of
1.53 and 1.66, respectively, indicating that the \sxr\ luminosity increases
faster than linearly with the 12\um\ luminosity, approximately as $L_{SX}
\propto L_{12\mu{m}}^{1.5}$ (we note that the scatter is large enough that one
of these luminosities can only be used to predict the other to within a factor
of about $\pm$5, if the Seyfert type is known). This trend is in the opposite
direction of what one would expect if a significant Malmquist bias were present
in this plot, as was shown to be the case for early reports that the X--ray
luminosity scaled less than linearly with optical luminosity (Chanan
1983). Since the 12\um\ luminosity is closely proportional to bolometric
luminosity for both Seyfert~1s and~2s (Spinoglio \& Malkan 1989; Spinoglio et
al. 1995), this relation shows either (1) that more bolometrically luminous
objects are less absorbed (intrinsically) in the \sxrs\ (Reichert et al. 1985),
or (2) that they are relatively more efficient at producing \sxrs\ in the first
place. Since the 12\um\ luminosity represents both nuclear and galactic
emission, the latter relation could simply imply that the more luminous objects
are more nuclear--dominated.


However, we note that a plot of the \sxr\ flux versus the 12\um\ flux does not
show a significant correlation for either Seyfert type.  It is well known that,
when no true correlation exists, as may be indicated by pure scatter in a
flux--flux diagram, the effect of redshift may introduce a spurious correlation
in the luminosity--luminosity diagram. However, the reverse scenario is also
possible: since luminosity is the physically meaningful parameter, it may be
that the luminosity--luminosity diagram reflects the true correlation, while
introducing the distance factor may smear out the correlation in a flux--flux
diagram.  Therefore, to determine whether there is truly a correlation, we have
also fit the data in the luminosity--luminosity diagram in a way that accounts
for the upper limits.  This can discriminate between the two cases just
mentioned, because, as Feigelson (1996) points out, ``A correlation found
between luminosities, when nondetections are taken into account, is a true
correlation. The flux--flux diagram will reveal the intrinsic relationship
between luminosities only under very limited circumstances (no censored points,
linear relationship)." (Also see Feigelson \& Berg 1983 and Feigelson 1992 for
further discussion of statistical methods for astronomical data with
nondetections.)

Because many of our \sxr\ luminosities, especially for the Seyfert~2s, are only
upper limits, we analyzed these data using ASURV Rev 1.2 (``Astronomy SURVival
Analysis" software; La Valley, Isobe, \& Feigelson 1992), which implements the
methods presented in Feigelson \& Nelson (1985) and Isobe, Feigelson, \& Nelson
(1986). ASURV provides statistical tests for the presence of a linear
correlation between two variables when either one (in our case, the RASS
luminosities) is heavily censored (e.g., mostly upper limits). In general, this
is done by making certain assumptions which allow one to estimate the
distribution function of the censored data (e.g., by maximum--likelihood
techniques), incorporating the information supplied by the detected points.
Using these methods, the generalized Kendall's tau correlation coefficient
(Brown, Hollander, \& Korwar 1974) indicates a 99.99\% probability that a
correlation is present for both the Seyfert~1s and the Seyfert~2s. The binned
two--dimensional Kaplan--Meier distribution and associated linear regression
coefficients of Schmitt (1985) were used to determine the linear best--fits,
shown as the long--dashed lines, which have slopes of $1.46\pm0.14$ for the
Seyfert~1s and for the $1.21\pm0.07$ Seyfert~2s. The Seyfert~1 fit has the same
slope, and is shifted down slightly, compared to the fit for detections only.
This is as expected, since only 18 of the 71 Seyfert~1s in this plot are not
detected in both parameters. However, the Seyfert~2 fit is lower, as expected
since most (40 of 64) points are censored in at least one variable. The ASURV
fit the the Seyfert~2s is also flatter than the fit to the detections only, but
still has a slope significantly greater than one. Thus, in each of these
relations, \lsx\ rises greater than linearly with 12\um\ luminosity, with
either the same slope for Seyfert~1s and~2s, or a slightly flatter slope for
the Seyfert~2s.  Almost the same relation is derived by Barcons et al. (1995),
who find $L_{5~keV}\propto L_{12}^{1.3}$ for Seyferts in the 12\um\ sample
(both types), with the Seyfert~1s having relatively $\sim7$ times the X--ray
luminosity as Seyfert~2s as compared to the 12\um\ luminosity.

We further note that the actual relation for Seyfert~2s (i.e. that which would
be obtained if we knew the \sxr\ luminosities for all Seyfert~2s in the 12\um\
sample) may fall even lower. This is because the survival--analysis routines
used assume that both the detected and undetected objects were drawn from the
same homogeneous sample, which can overestimate the importance of the censored
data in determining the fits. This was seen, for example, in RMS, where we used
ASURV to fit a line to a plot of $L_{6cm}$ vs. $L_{60\mu m}$, for Seyferts in
the 12\um\ sample, including many upper limits to the 6~cm luminosity. When we
later (Rush, Malkan, \& Edelson 1996) obtained 6~cm detections, we found the
actual fit to be parallel, but lower, i.e. 0.3 magnitudes less $\log L_{6cm}$
for a given $\log L_{60\mu m}$.

Although the slopes may be similar for each Seyfert type, we see the two fits
to be clearly shifted from each other, with the \sxr\ luminosity of Seyfert~2s
ranging from 1.1 to 1.3 orders of magnitude lower than that of Seyfert~1s for a
given 12\um\ luminosity, as the latter ranges from 42.5 to 44.5 (according to
the ASURV fits). In fact, {\it all\/} objects with $\log
L_{SX}/L_{12\mu{m}}>-1$ are Seyfert~1s, values from --2 to --1 include both~1s
and~2s, while most objects below --2 are Seyfert~2s (the dotted lines indicate
$\log L_{SX}/L_{12\mu{m}}=-1$ and --2). Similar results were obtained by Green,
Anderson, \& Ward (1992) who showed that galaxies with $\log
L_{SX}/L_{60\mu{m}}>-2$ (using Einstein 0.5---4.5~keV fluxes) are almost
certain to show broad--line optical emission. This is quantitatively consistent
with our result, assuming typical colors for Seyferts of $\log
F_{60}/F_{12}=0.5-1.0$ (RMS) and assuming an Einstein \pin\ of 1.9 (typical for
our sample, as estimated from the 20 Seyferts in our sample with measured IPC
photon indices in Kruper et al. 1990).


\subsection{Infrared--Luminous Non--Seyferts} \label{results_ILNS}

The X--ray--to--IR relation of IR--luminous non--Seyfert galaxies in our sample
is distinct from that of Seyfert~1s, but is distributed roughly the same as the
the Seyfert~2s. This is shown in Figure~4{\it{b\/}} (and was also evident in
Figure~1{\it{b}}), where we plot the \sxr\ versus 12\um\ luminosities (the same
parameters as in Figure~4{\it{a\/}}), showing only the detections for clarity,
and also plotting the IR--luminous non--Seyferts in the 12\um\ sample (as
stars). In this plot, the six IR--luminous non--Seyferts detected in our sample
have values of $\log L_{X}/L_{IR}$ between --2.8 and --1.8. This
Seyfert~2--like ratio indicates that these objects may harbor active nuclei
which would be obscured by dust, or, conversely, that the 12\um\ and \sxr\
flux from some Seyfert~2s may be from starbursts. Boller et al. (1996a), for
example, quantify this by calculating the maximum $L_X$ that could be emitted
from various stellar X--ray sources (O--stars, galactic superwinds, X--ray
binaries, or SNRs). They calculate that a value of $\log L_{X}/L_{IR}$ of
$\sim10^{-2}$ is just around the maximum ratio which can be produced by SNRs,
and therefore higher values require that additional nuclear emission must be
present. Such objects would be observed as AGN (probably Seyfert~2s) only upon
inspection of high--SNR, high--resolution spectroscopy, or perhaps requiring IR
spectroscopy if the AGN is completely obscured by the dust in the optical. That
this may indeed be the case is highlighted by the fact that three of the
Seyfert~2s in this plot (individually labeled) were actually included in our
IR--luminous class at the start of this project, but have since been discovered
to be AGN upon inspection of optical spectra, and {\it each\/} of these objects
was studied spectroscopically specifically because their high \sxr\
luminosities made them interesting Seyfert candidates. For example, realizing
the power of strong X--rays as an indicator of buried AGN, we proposed a
pointed ROSAT PSPC observation of the most X--ray--luminous non--Seyfert in the
sample, MCG--4--2--18 ($L_X=42.9$ and $\log L_{X}/L_{IR}=-1.2$). That
observation was eventually awarded to Ebeling who measured a count rate and fit
a power--law to the PSPC spectrum which are consistent with our RASS data,
indicating no variability in the 1.5--2 years between those observations
Furthermore, optical spectroscopy of that object now identifies it as a Seyfert
galaxy (H.~Ebeling, private comm.). Two similarly X--ray luminous galaxies
(also labeled in the figure---NGC~3147 and NGC~3822) were recently identified
as Seyfert~2s from their optical spectra by Moran, Halpern, \& Helfand (1994),
who also were interested in these galaxies because of their notable X--ray
strength (as reported in Boller et al.---1992, 1993). Given these
identifications, the remaining supposed non--Seyferts, especially the most
X--ray--luminous ones (e.g., N6240 and ESO~286-IG19, which have $L_X=42.3$ and
42.7, respectively), are likely candidates to have an obscured active nucleus.




\section{Interpretation and the Unified Model of Seyferts} \label{interp} 

One explanation for the observed differences between Seyfert~1s and Seyfert~2s,
with regards to both their \sxr\ and multiwavelength properties, is the
obscuring--torus model for Seyferts (Antonucci \& Miller 1985). In this model,
the two types of Seyferts are intrinsically similar objects\footnote{As
mentioned in \SS{targets_calcs}, we use the terms ``intrinsic'' and
``observed'' to distinguish bewteen nuclear emission from the target galaxy and
that which has possibly been absorbed, obscured, and/or scattered by \nh\ or
other material in the target galaxy. Both terms are used to refer to the flux
or luminosity as would be measured before being absorbed by \nh\ in our
Galaxy.}, with a molecular torus (or similar anisotropic obscuring material)
surrounding the innermost regions. This torus shields the central engine and
broad--line region gas from our direct view in edge--on objects (i.e.,
Seyfert~2s) but not in face--on objects (i.e., Seyfert~1s). However, the energy
emitted in the inner region can be scattered into our \los\ by a cloud of
electrons above and/or below the torus. Thus, if the \sxrs\ originate from
radii within the torus, then the amount we observe from Seyfert~2s is only the
fraction which is intercepted and scattered towards our \los\ as it
escapes. Since the cross--section to electron scattering is
wavelength--independent, this would result in Seyfert~2s having the same
distribution in \gam\ as Seyfert~1s, as is seen in Figures~1 and~2 above.
However the scattering would only transmit a small fraction of the \sxrs\
(proportional to $\tau_{es}\times\frac{\Omega_c}{4\pi}$, where $\Omega_c$ is
the solid angle of the scattering region seen by the nucleus---Miller,
Goodrich, \& Matthews 1991), so a given Seyfert~2 would also have a lower
observed \sxr\ luminosity than an intrinsically similar Seyfert~1, also as seen
in Figure~1{\it{b}}.

Furthermore, according to this model, if the 12\um\ emission arises totally or
primarily from regions at large radii from the center, i.e. mostly outside of
the torus, or if the optical depth of the torus to such photons is low enough
to allow them mostly to pass through, then it would be mainly isotropic and
would be the same in Seyfert~1s and~2s. This is also consistent with what we
see in Figures~4 above, namely that Seyfert~1s and~2s span a similar range in
mid--infrared luminosity, but that for a given value of this low--frequency
luminosity, the \sxr\ emission from Seyfert~2s is one and a half orders of
magnitude lower. Combining these effects, we have a picture of Seyfert galaxies
in which face--on objects have a certain intrinsic range in the \sxr\ and
12\um\ luminosities and \sxr\ slopes, and are optically identified as
Seyfert~1s.  When viewed at high inclinations, the torus obscures the innermost
regions, suppressing the observed \sxr\ luminosity, and shielding the
broad--line region to cause the objects to be optically identified as
Seyfert~2s, but without changing either the \sxr\ slope or lower--frequency
luminosity. Thus, the unified model is consistent with all of our data if, as
predicted by the difference between the \sxr\ luminosities of Seyfert~1s
and~2s, the scattering efficiency is typically $\sim$1--5\%, with not too much
variation from one galaxy to another.

An alternative explanation is that the \sxrs\ we are observing from Seyfert~2s
are not scattered, but suffer higher amounts of internal absorption than those
from Seyfert~1s (Lawrence \& Elvis 1982). Either case would have the effect of
lowering the observed \sxr\ flux, while not changing the lower--frequency flux
(which would not be absorbed). However, we prefer the scattering explanation
for two reasons: (a) there is little independent evidence for significant
amounts of internal absorption in the soft X--rays (except in a few individual
cases, e.g., NGC~1068) in papers which analyze ROSAT pointed observations of
Seyfert~2s (e.g., Turner, Urry, \& Mushotzky 1993; Rush \& Malkan 1996); and
(b) internal absorption would flatten the observed \pin\ in Seyfert~2s and thus
the distribution of \gam\ would be significantly different between the two
types of objects, contrary to what we see in Figures~1 and~2. However, we note
that the results of the K--S test on \gam\ may be fortuitous, given the small
number of Seyfert~2s used, and the possible selection effects which could
result if the faint, undetected objects have a have a different distribution of
\gam\ than does the sample as a whole. A variation of this possibility is that
the nuclear \sxrs\ from many (if not most) Seyfert~2s are {\it totally\/}
internally absorbed, and what we observe is extended emission only, e.g., as
discussed above when comparing the Seyfert~2s to IR--luminous
non--Seyferts. However, this would probably also predict a different
distribution of \gam\ between the two Seyfert types.

Another alternative to the scattering picture, which has already been used to
explain some objects, is that there may be multiple components measured in the
ROSAT passband in Seyferts. For example, recent studies of Einstein--selected
``ultra--soft X--ray emitting AGN" (e.g., Puchnarewicz et al. 1992; C\'ordova
et al. 1992; Thompson 1994) have identified an entire class of objects with
clearly distinguishable soft X--ray components below 0.5~keV, possibly
corresponding to a $\sim10$~eV blackbody. This sample contains a large number
of narrow--line objects (e.g., Seyfert~2s, narrow--line X--ray galaxies,
narrow--line Seyfert~1s) as compared to other X--ray selected samples (which
they suggest may be caused by face on BLRs, or BLRs far from the central
source), but is otherwise similar to such samples (e.g., in optical luminosity,
and optical Fe~II emission).  Higher--energy components to the \sxrs, such as
iron and/or oxygen emission lines around 0.7--0.8~keV have also been suggested
as a likely source for a soft excess in some individual Seyferts, particularly
in Seyfert~2s (e.g., Turner et al. 1993; Rush \& Malkan 1996). However, the
distributions of gamma shown in Figure~1 makes this seem unlikely as a {\it
universal\/} explation for the differences between~1s and~2s.  For example, to
account for the higher X--ray luminosities of Seyfert~1s, one might propose
that their ROSAT flux is explained by an underlying power--law plus a soft or
ultra--soft excess, but that we only measure the latter in Seyfert~2s (either
because the power--law is component non--existent or because it is simply not
observed, perhaps being obscured by the torus). But this explanation would
require Seyfert~2s to have systematically steeper ROSAT spectra, which is not
observed in our samples.

Finally, we note that a combination of these explanations is possible. If, for
example, some objects in our sample have more internal absorption, this would
tend to flatten the observed slope. On the other hand, if these same objects
also have a soft or ultra--soft excess (which could originate in the same
region producing the extra absorption), this factor would steepen the observed
slope.  These two factors could cancel each other, leaving the measured \pin\
roughly unchanged, as compared to the situation in which neither factor
exists. This would be consistent with what we find in Figure~1, namely
distributions of \gam\ which are not significantly different between Seyfert~1s
and Seyfert~2s.

\section{X--Ray Luminosity Functions} \label{xlfs} 

An X--ray luminosity function (XLF) has been derived for the Seyferts in the
12\um\ sample (type~1 and type~2, including the few NLS1s and NLXGs).  We have
used the $V/V_{max}$ method (Schmidt 1968; Schmidt \& Green 1983),
$$\Phi={4\pi\over \Omega f\Delta{L}}\sum{1\over V_{max}},$$ where Omega is the
area of sky surveyed, $f$ is the fraction of objects in our sample observed by
the RASS (i.e., located in the area of sky surveyed by ROSAT, whether detected
or not), and $\Delta{L}$ is the bin size.  We have computed $V_{max}$
individually for each galaxy in the sample. In doing so, we have followed the
method of Edelson (1987) for calculating the luminosity function of a sample at
a wavelength other than the wavelength at which the sample was defined (that
work calculated a radio LF for an optically--selected sample). We thus use
$$V_{max} = \mbox{min} (V_{max,survey} V_{max,RASS}),$$ which represents the
maximum volume of space accessible by an object detected at the survey
wavelength (mid--IR for the 12\um\ sample) and at X--ray wavelengths. This is
equivalent to deriving the XLF from the 12\um\ luminosity function and from the
bivariate IR--X--ray luminosity distribution function (Elvis et al. 1978; Meurs
\& Wilson 1984). We stress, however, that this luminosity function is only that
for the 12\um\ sample, with certain IR--X--ray selection effects (and can at
best be considered as a lower limit to the true X--ray space density of
Seyferts). In particular, the most extreme objects (i.e., those with 12\um\
fluxes below the 12\um\ survey limit, yet \sxr\ count rates above the RASS
detection limit) will be excluded, having not been included in the sample in
the first place, even though they would have been detected in the \sxrs.
(Fortunately, the IR--X--ray relation shown in Figures~4 suggests that such
objects would be rare.)


The XLF for the Seyferts in the 12\um\ sample is given in Table~2 and plotted
in Figure~5. The errorbars represent the 90\% confidence interval, based on
Poisson statistics, calculated using the equations from Gehrels (1986) which
are accurate for even very small numbers of data points.  The solid line shows
a least--square fit with a single power--law (the points are weighted by the
number of objects they represent, as noted by the numbers under the solid
squares, hence the line looks higher than it would if each point were weighed
evenly). This fit shows the XLF to be a power--law with a well-defined slope of
--1.75 and a correlation coefficient of $r=-0.98$. An F--test indicates that no
significant improvement to the fit is obtained when using a broken power--law
over a single power--law.


To compare our ROSAT \sxr\ XLF to that from the HEAO1 \hxr\ sample of
Piccinotti et al. (1982; hereafter P82), we converted their values (both data
points and fit) to the ROSAT passband. We assumed a spetrum which has a
0.1---2.0~keV photon index of 2.3 (the median from our sample), a
2.0---10.0~keV photon index of 1.7 (typical for \hxr\ samples as mentioned
above), with these two components normalized to the same level at 2.0~keV. (Any
reasonable variations in the slopes chosen to convert their XLF contribute
negligible differences to the characteristics of this plot.) We also converted
their XLF to the units of $H_0=75$~\kmsmpc\ used in this paper. The resultant
XLF from P82 is shown by the dotted line and open squares.  The error bars also
represent 90\% confidence interval, calculated in the same way as those for the
12\um\ sample (using Gehrels 1986; each of their bins includes only 3 objects,
and thus all have the same sized error bars). Although the P82 XLF is steeper
(slope = --2.5), the two XLFs agree within the errors over the whole range in
$L_X$ spanned by both samples, as shown by the shorter solid line, which shows
the fit to our sample above $\log L_X = 42.5$ (slope = --2.12, $r=-0.96$).



The close agreement of the 12\um--sample XLF with that calculated from the {\it
hard--X--ray selected\/} sample of P82 is further support for the argument that
12\um\ selection is relatively unbiased (and that our bivariate XLF is a close
approximation to the purely X--ray XLF for AGN). In other words, there are few,
if any, objects which are particularly X--ray--strong but weak at 12\um. This
is because the fraction of bolometric luminosity emitted at 12\um\ is
insensitive to the overall spectral energy distribution of a given AGN
(Spinoglio \& Malkan 1989; RMS; Spinoglio et al. 1995), and because the scatter
in the IR--X--ray relation for a given Seyfert type is small compared to the
range of luminosities spanned.  This result is also confirmed in the hard
X--rays by Barcons et al. (1995), who compare the 12\um\ sample to a \hxr\
selected sample, and find that most of the observed 2---10~keV X--ray
luminosity function over 10$^{42-46}$~erg~s$^{-1}$ can be accounted for by
12\um--emitting AGNs.  Conversely, the 12\um\ LF for Seyferts calculated from
an X--ray--selected sample would probably be too low, since such samples miss
those Seyferts which are weak in X--rays.  We would also expect a
60\um--selected sample, for example, to yield a lower XLF, because objects with
higher X--ray luminosities are more likely to have lower far--infrared
luminosities, being low in dust content (e.g., the Markarian Seyferts).



\section{Summary} \label{summary} 

We have analyzed the results of ROSAT All--Sky Survey observations of Seyfert
and IR--luminous galaxies from the Extended 12\um\ Galaxy Sample and the
optically--selected CfA Sample. For those objects with enough counts to be
fitted to an absorbed power--law model, the average \pin\ is \gam=2.26$\pm$0.11
for 43 Seyfert~1s and \gam=2.45$\pm$0.18 for 10 Seyfert~2s.

Any internal absorption has the affect of flattening the measured \pin\ and
decreasing the observed luminosity, although this affect is likely significant
in only the most obscured objects (we describe four Seyfert~1.5s for which this
internal absorption appears to be significant). Absorption by \nh\ in our
Galaxy has likely caused some Seyferts to go undetected by ROSAT (we estimate
as many as close to 5--10 in our sample), especially where
$N_{H,gal}\gapprox4\times10^{20}\mbox{cm}^{-2}$ along our \los.

Although Seyfert~1s and~2s have similar distributions and median values of
\gam, the median \sxr\ luminosity for Seyfert~1s is an order of magnitude
higher than that for Seyfert~2s.  We find correlations of \lsx\ with the 12\um\
luminosity, with similar slopes for both types of Seyferts, and with Seyfert~1s
having relatively more \sxr\ emission by 1---2 orders of magnitude, relative to
the mid--infrared emission.  These observations are consistent with the
obscuring--torus model for Seyferts if the soft X--rays are scattered into our
\los\ in Seyfert~2s (with a scattering efficiency typically of $\sim$1--5\%),
which would decrease the observed \sxr\ flux while maintaining the same \pin\
as compared to observing the same object face--on as a Seyfert~1.  Thus,
observing direct versus scattered emission could be the primary difference
between Seyfert~1s and~2s, with internal absorption modifying the results
slightly.  Alternative explanations require fortuitous agreement of Seyfert~1
and~2 \sxr\ spectral shapes.

Infrared--luminous non--Seyferts are shown to resemble the more Seyfert~2s in
their distribution of \sxr\ luminosities ($\log L_X\sim 41.4-42.7$)
and X--ray--to--IR colors. That several of these objects may indeed harbor
obscured active nuclei is implied by the fact that three similar objects in our
sample were only recently identified to be Seyfert~2s based on optical spectra.

The \sxr\ luminosity function of the 12\um\ sample is well described by a
single power--law with a slope of --1.75. This XLF is generally consistent with
that of the \hxr\ sample of Piccinotti (1982).  Barcons et al. (1995) made a
similar analysis of the 12\um\ sample in the \hxrs, also finding a high
contribution of 12\um--emitting AGN to the XLF, showing that 12\um\ detection
does not miss many \hxr\ selected AGN.  They also determined values for the
slope of the X--ray--to--IR relation, and the ratio of Seyfert~1 to Seyfert~2
average \sxr\ luminosities which are consistent with our results.


\acknowledgements

We thank the staff at MPE for assistance in processing the RASS data. We thank
the anonymous referee for a very thorough reading of the manuscript and many
helpful comments. This work was supported in part by NASA grants NAG~5--1358
and NAG~5--1719.  This research has made use of the NASA/IPAC Extragalactic
Database (NED) which is operated by the Jet Propulsion Laboratory, California
Institute of Technology, under contract with the National Aeronautics and Space
Administration.



\clearpage \onecolumn 

\begin{table} 
\begin{center} 
TABLE 2\\
\vspace{0.4cm}
{\sc Seyfert Galaxy X--Ray Luminosity Function}\\
\vspace{0.3cm}
\begin{tabular}{ccc}\hline\hline
\mc{1}{c}{$\log L_{SX}^a$}&
\mc{1}{c}{$\log\Phi$}&
\mc{1}{c}{}\\
\mc{1}{c}{(ergs~s$^{-1}$)}&
\mc{1}{c}{($\Delta L_{44}^{-1}$ Mpc$^{-3}$)}&
\mc{1}{c}{$N$}\\
\hline
   40.25 &    0.32 &   3 \\
   40.75 &   -0.62 &   7 \\
   41.25 &   -1.01 &   5 \\
   41.75 &   -2.02 &  10 \\
   42.25 &   -3.24 &  11 \\
   42.75 &   -3.33 &  10 \\
   43.25 &   -5.05 &   5 \\
   43.75 &   -4.84 &   7 \\
   44.25 &   -7.17 &   2 \\
   44.75 &   -7.71 &   3 \\
   45.75 &  -11.43 &   1 \\
   46.25 &   -9.49 &   1 \\
\hline
\end{tabular}
\end{center}
\begin{center}
\parbox[c]{5cm}{\noindent
\hspace*{.3cm}$^a$~Bin center, for bins of width 0.5 in $\log L_{SX}$}
\end{center}
\end{table}

\clearpage 


\clearpage 

\centerline{\bf FIGURE LEGENDS}

\noindent {\bf Figure~1} --- Differential histograms of {\it{a}}: the \PIN,
\gam, and {\it{b}}: the \sxr\ luminosity. Densely--hatched squares are
Seyfert~1s, sparsely--hatched squares are Seyfert~2s, and open squares are
IR--luminous non--Seyferts. For the Seyferts: horizontal, diagonal, and
vertical lines represent objects in the 12\um\ sample, both samples, and the
CfA sample, respectively.

\noindent {\bf Figure~2} --- \gam\ versus the unabsorbed observed flux at
1~keV.  In this and all following plots (unless otherwise noted): Filled
symbols are Seyfert~1s, open symbols are Seyfert~2s, and stars are IR--luminous
non--Seyferts. For the Seyferts: squares are in the 12\um\ sample only,
triangles are in the CfA sample only, and circles are in both samples.  The two
objects in the upper--right portion of this diagram have uncertain values for
\gam, as explained in the text.

\noindent {\bf Figure~3} --- Flux at 1~keV versus \nhg. The box in the
lower--left indicates the region of low Galactic obscuration through which even
faint Seyferts can be detected.

\noindent {\bf Figure~4} --- \Sxr\ Luminosity versus the monochromatic 12\um\
luminosity. {\it{a}}: Seyfert~1s and~2s. The higher and lower solid lines are
fits to detected Seyfert~1s and and detected Seyfert~2s, respectively. The
higher and lower long--dash lines are fits to Seyfert~1s and Seyfert~2s,
respectively, accounting for upper limits using the ASURV software.  Dotted
lines denote $\log L_{SX}/L_{12\mu{m}}=-1$ and $\log L_{SX}/L_{12\mu{m}}=-2$.
{\it{b}}: detections only, for Seyfert~1s, Seyfert~2s, and IR--luminous
non--Seyferts. Three recently identified Seyfert~2s are individually labeled.

\noindent {\bf Figure~5} --- X--ray luminosity function of Seyfert~1 galaxies
in the 12\um\ sample, fit to a straight--line power--law (long solid line and
filled squares), with each point weighted by the number of galaxies it
represents (as noted by the numbers under the squares). Shown for comparison
are the data and fits from Piccinotti et al. (1982; dotted line and open
squares), converted to lower energies and an $H_0$ of 75, as described in the
text. The short solid line is the fit to the 12\um\ XLF in the range spanned by
the Piccinotti data.


\begin{references}
\rf{Antonucci, R.R.J. \& Miller, J.S. 1985, ApJ 297, 621}
\rf{Awaki, H. 1993, in Frontiers of X-ray Astronomy, ed. Y. Tanaka \& K. Koyama
 (Tokyo: Universal Academy Press, Inc.), 573}
\rf{Bade, N., Fink, H.H., Engels, D., Voges, W., Hagen, H.--J., Wisotzki, L.,
 \& Reimers, D. 1995, A\&AS 110, 469.}
\rf{Bade, N. \& Schaeidt, S. 1994, in Multiwavelength Continuum Emission of 
 AGN, IAU Symposium 159, ed. T.J.--L. Courvoisier \& A. Blecha  (Dordrecht: 
 Kluwer Academic Publishers), 365.}
\rf{Barcons, X., Franceschini, A., De~Zotti, G., Danese, L., \& Miyaji, T.
 1995, ApJ 455, 480}
\rf{Boller, Th. et al. 1996a, A\&A, submitted}
\rf{Boller, Th., Meurs, E.J.A., Brinkman, W., Fink, H., Zimmerman, U., \&
 Adorf, H.--M. 1992, A\&A 261, 57}
\rf{Boller, Th., Tr\"umper, J., Molendi, S., Fink, H., Schaeidt, S., Caulet,
 A., \& Dennefeld, M. 1993, A\&A 279, 53} 
\rf{Brinkmann, W. 1992, in ``Physics of Active Galactic Nuclei", Eds. Duschl,
 W.J. \& Wagner, S.J. (Berlin: Springer--Verlag), 19}
\rf{Brinkmann, W., Siebert, J., Reich, W., F\"urst, E., Reich, P., Voges, W.,
 Tr\"umper, J., \& Wielebinski, R. 1995, A\&AS 109, 147}
\rf{Brown, B.J. Jr., Hollander, M., \& Korwar, R. M. 1974, in Reliability and
 Biometry, eds. F. Proschan and R. J. Serfling (SIAM: Philadelphia) p.327} 
\rf{Chanan, G.A. 1983, ApJ 275, 45}
\rf{C\'ordova, F.A., Kartje, J.F., Thompson, R.J.J., Mason, K.O., Puchnarewicz,
 E.M., \& Harnden, F.R.J. 1992, ApJS 81, 661}
\rf{Dickey, J.M., \& Lockman, F.J. 1990, ARA\&A 28, 215}
\rf{Edelson, R.A. 1987, ApJ 313, 651}
\rf{Elvis, M., Maccacaro, T., Wilson, A.S., Ward, M.J., Penston, M.V., Fosbury,
 R.A.E., \& Perola, G.C. 1978, MNRAS 183, 129}
\rf{Feigelson, E. D. 1992, in ``Statistical Challenges in Modern Astronomy'',
 Eds. Feigelson, E.~D. and Babu, G.~J. (Springer--Verlag), 221}
\rf{Feigelson, E.D. 1996, in a note titled ``Flux vs. Flux or Luminosity vs.
 Luminosity'', from the World Wide Web resource of the SCCA (Statistical 
 Consulting Center for Astronomy) operated at the Department of Statistics, 
 Penn State University, M.~G.~Akritas (Director)}
\rf{Feigelson, E. D. \& Berg, C. J. 1983, AJ 269, 400}
\rf{Feigelson, E. D. \& Nelson, P. I. 1985, ApJ 293, 192} 
\rf{Gehrels, N. 1986, ApJ 303, 336}
\rf{Green, P.J., Anderson, S.F., \& Ward, M.J. 1992, MNRAS 254, 30}
\rf{Hasinger, G., Schmidt, M., \& Tr\"umper, J., 1991, A\&A 246, L2.} 
\rf{Heiles, C. \& Cleary, M.N. 1979, AuJPA 47, 1}
\rf{Huchra, J. \& Burg, R. 1992, ApJ 393, 90}
\rf{Isobe, T. Feigelson, E. \& Nelson, P. 1986, ApJ 306, 490} 
\rf{Kriss, G.A., Canizares, C.R., \& Ricker, G.R. 1980, ApJ 242, 492}
\rf{Kruper, J.S., Urry, C.M., \& Canizares, C.R. 1990, ApJS 74, 347}
\rf{La Valley, M.P., Isobe, T. \& Feigelson, E.D. 1992, BAAS 24, 839} 
\rf{Lawrence, A. \& Elvis, M. 1982, ApJ 256, 410}
\rf{Meurs, E.J.A. \& Wilson, A.S. 1984, A\&A 136, 206}
\rf{Miller, J.S., Goodrich, R.W., \& Matthews, W.G. 1991, ApJ 378, 47}
\rf{Moran, E.C., Halpern, J.P., Bothum, G.D., \& Becker, R.H. 1992, AJ 104,
 990}
\rf{Moran, E.C., Halpern, J.P., \& Helfand, D.J. 1994, ApJL 433, L65}
\rf{Mushotzky, R.F., Done, D. \& Pounds, K.A. 1993, ARA\&A 31, 717}
\rf{Osterbrock, D.E. \& Martel, A. 1993, ApJ 414, 552}
\rf{Persic, M. et al. 1989, ApJ 344, 125}
\rf{Piccinotti, G., Mushotzky, R.E., Boldt, E.A., Holt, S.S., Marshall, E.E.,
 Serlemitsos, P.J., \& Shafer, R.A. 1982 ApJ 253, 485 (P82)}
\rf{Puchnarewicz, E.M., Mason, K.O., C\'ordova, F.A., Kartje, J.,
 Branduardi--Raymont, G., Mittaz, J.P.D., Murdin, P.G., \& Allington--Smith, J.
 1992, MNRAS 256, 589}
\rf{Reichert, G.A., Mushotzky, R.F., Holt, S.S., \& Petre, R. 1985, ApJ 296,
 69}
\rf{Rush, B., Malkan, M.A., \& Edelson 1996, ApJ, submitted}
\rf{Rush, B., Malkan, M.A., \& Spinoglio, L. 1993, ApJ 89, 1 (RMS)}
\rf{Rush, B. \& Malkan, M.A. 1996, ApJ, 456, 466}
\rf{Schmidt, M. 1968, ApJ 151, 393}
\rf{Schmidt, M. \& Green, R.F. 1983, ApJ 269, 352}
\rf{Schmitt, J.H.M.M 1985, ApJ 293, 178} 
\rf{Spinoglio, L. Malkan, M.A., Rush, B., Carrasco, L., \& Recillas--Cruz, E.
 1995, ApJ, in press., Vol. 453}
\rf{Spinoglio, L. \& Malkan, M.A. 1989, ApJ 342, 83}
\rf{Stark, A.A., Gammie, C.F., Wilson, R.W., Bally, J., Link, R.A., Heiles, C.,
 \& Hurwitz, M. 1992, ApJS 79, 77}
\rf{Thompson, R.J. 1994, PASP 106, 1222}
\rf{Turner, T.J., Urry, C.M.,   \&  Mushotzky, R.F. 1993, ApJ 418, 653}
\rf{Voges, W. 1992, in Proceedings of the ISY Conference ``Space Science'', ESA
 ISY--3, ESA Publications, 9.}
\rf{Ward, M.J., Hughes, D.H., Dunlop, J.S., \& Appleton, P.N. 1994, in
 Multiwavelength Continuum Emission of AGN, IAU Symposium 159, ed. T.J.--L.
 Courvoisier \& A. Blecha  (Dordrecht: Kluwer Academic Publishers), 311} 
\rf{Weedman, D.W., Feldman, F.R., Balzano, V.A., Ramsey, L.W., Sramek,
 R.A., \& Wu, C.--C. 1981, ApJ 248, 105} 
\rf{Wilkes, B.J. \& Elvis, M. 1987, ApJ 323, 243}
\end{references}
\end{document}